\documentclass[journal]{IEEEtran}
\usepackage{graphicx}
\usepackage{subfig}

\begin{document}
\title{Construction and operation of a Double Phase LAr Large Electron Multiplier Time Projection Chamber}

\author{A.~Badertscher,
        L.~Knecht,
        M.~Laffranchi,
	A.~Marchionni,
	G.~Natterer,
	P.~Otiougova,
	F.~Resnati,
	A.~Rubbia
\thanks{Manuscript received November 14, 2008. This work was supported by Swiss National Foundation and ETH Zurich.}\\
Institute for Particle Physics, ETH Zurich, 8093 Zurich, Switzerland%
}

\maketitle
\pagestyle{empty}
\thispagestyle{empty}

\begin{abstract}
We successfully operated a novel kind of LAr Time
Projection Chamber based on a Large Electron Multiplier~(LEM) readout
system. The prototype, of about 3 liters active volume, is operated in
liquid-vapour (double) phase pure Ar. The ionization electrons, after
drifting in the LAr volume, are extracted by a set of grids into the gas phase and
driven into the holes of a double stage LEM, where charge amplification occurs.
Each LEM is a thick macroscopic hole multiplier of 10x10~cm$^2$
manufactured with standard PCB techniques. The electrons signal is
readout via two orthogonal coordinates, one using the induced
signal on the segmented upper electrode of the LEM itself and the
other by collecting the electrons on a segmented anode. Custom-made
preamplifiers have been especially developed for this purpose.
Cosmic ray tracks have been successfully observed in pure gas at room temperature and
in double phase Ar operation. We believe
that this proof of principle represents an important milestone in the
realization of very large, long drift (cost-effective) LAr detectors
for next generation neutrino physics and proton decay experiments, as
well as for direct search of Dark Matter with imaging devices.
\end{abstract}

\section{Introduction}
\IEEEPARstart{I}{n} this paper we will describe the construction and the results of a small liquid argon LEM Time Projection Chamber which 
we believe to be scalable to large size LAr detectors~\cite{Rubbia04}. Differently from ICARUS LAr TPC~\cite{ICARUS}, the LAr~LEM-TPC is operated in double 
phase (liquid-vapour) pure Ar to allow signal amplification in the gas phase. The ionization electrons, after drifting in the LAr volume, 
are extracted by a set of grids into the gas phase and driven into the holes of a double stage Large Electron Multiplier~(LEM) device, where charge amplification occurs. 
Each LEM is a thick macroscopic multiplier manufactured with 
standard PCB techniques, directly extrapolated from the more delicate GEM detectors~\cite{Sauli97}. 
The use of the LEM is motivated by the following facts:
a wide range of gains is achievable, from $\sim$10 to $\sim$10$^3$; the gain is easily adjustable to a large spectrum of  physics requirements;
it is a sturdy detector capable of cryogenic operation;
large surfaces can be covered as a collection of individual $\sim$1x1~m$^2$ pieces (the largest size 
presently manufactured);
as shown for the GEM detectors~\cite{Sauli99}, LEMs can be operated in pure Ar gas, as required in double-phase 
operation.
LAr~LEM-TPC detectors have improved imaging capabilities and could be compatible with very long drift paths.
When built with radio-pure materials and operated with gains as large as $\sim$10$^3$, they could also be used for nuclear recoil detection~\cite{ArDM}.
The LEM construction and its working principle are described in sections~\ref{sec:holes},~\ref{sec:LEM}, while the construction details of the LAr~LEM-TPC
are addressed in section~\ref{sec:LEM-TPC}. A specific effort was devoted to the readout components, in particular 
to the choice of passive elements compatible with cryogenic conditions, and
to the design and construction of custom-made preamplifiers
as  reported in section~\ref{sec:electronics}.
The detailed performance of a LEM-TPC in pure Ar gas at ambient temperature is described in section~\ref{sec:gas}.
Finally, the proof of operation 
of a double phase LAr~LEM-TPC as a tracking device is given in section~\ref{sec:liquid}.

\section{Electron avalanche multiplication in small holes}
\label{sec:holes}
The general technique of electron multiplication via avalanches in small holes
is attractive because (1)~the required high electric field can be naturally attained
inside the holes and (2)~the finite size of the holes effectively ensures 
a confinement of the electron avalanche, thereby reducing secondary 
effects in a medium without quencher. 

The gain ($G$) in a given uniform electric field of a parallel plate chamber  
at a given pressure is described by $G\equiv~e^{\alpha d}$ where 
$d$ is the gap thickness and  $\alpha$
 is the Townsend coefficient, which represents the number of electrons
 created per unit path length by an electron in the amplification region. 
The behavior of this coefficient 
with pressure and electric field can be approximated by the Rose and 
Korff law~\cite{RoseKorff}: $\alpha=A\rho~e^{-B\rho~/E}$
where  $E$ is the electric field, 
$\rho$ is the gas density, $A$ and $B$ are the parameters depending on the gas.

Electron multiplication
in holes has been investigated for a large number of applications.
The most extensively studied device is the  Gas Electron Multiplier~(GEM)~\cite{Sauli97}, 
made of
50--70~$\mu$m diameter holes etched in a 50~$\mu$m thick metalized
Kapton foil.  Stable operation has been shown with various gas mixtures
and very high gains.
An important step was the operation of the GEM in 
pure Ar at normal pressure and temperature~\cite{Sauli99}. 
Rather high gas gains were obtained, of the order of 1000, supporting  evidence
for the avalanche  confinement to the GEM micro-holes.


Operation of GEMs in an avalanche mode in pure Ar in double phase conditions has been
studied~\cite{Bondar06}, 
using triple-stage GEMs reaching gains of the order of 5000.

The successes of the GEMs triggered the concept of the Large Electron
Multiplier~(LEM) or THGEM (for a recent review see~\cite{Bondar08}), a coarser but more rigid
structure made with holes of the millimeter-size in a millimeter-thick
printed circuit board~(PCB).



In order to study the properties of the LEM and the possibility to
reach high gains in double phase, we have performed extensive R\&D on several prototypes~\cite{Polina_ETHthesis}:
a first single stage prototype demonstrated a stable operation in pure Ar 
at room temperature and pressure up to 3.5~bar with a gain of 800 per electron. 
Measurements were performed at high pressure because the density of Ar at 3.5~bar is roughly
equivalent to the expected density of the vapour at the temperature of 87~K. Simulations 
of the LEM operation were performed using the MAXWELL (field calculations) and 
MAGBOLTZ (particle tracking) programs. The results obtained were in good agreement 
with the experiment. The simulations showed that a double-stage LEM system is preferred
to reach gains of $\sim$10$^3$.
Double stage LEM configurations were tested in pure Ar at room
temperature, cryogenic temperature and in double phase conditions.
Tests in an Ar/CO$_2$ (90\%/10\%)
gas mixture were also performed to compare the results with those obtained in pure Ar.
The double-stage LEM system demonstrated a gain of $\sim$10$^3$ at a temperature of 87~K
and a pressure of $\sim$1~bar. The double-phase operation of the LEM proved the extraction
of the charge from the liquid to the gas phase.

\section{Large Electron Multiplier~(LEM)}
\label{sec:LEM}
We have built several LEM prototypes using standard PCB techniques from different manufacturers.
Double-sided copper-clad (16~$\mu$m layer) FR4 plates with thicknesses ranging from 0.8~mm to 1.6~mm
are drilled with a regular pattern of 500~$\mu$m diameter holes  at a relative distance of 800~$\mu$m. 
We report here on the 1.6~mm thick LEM.

By applying a potential difference on the two faces of the PCB an intense electric field inside the holes is produced.
\begin{figure}[htb]
\centering
\includegraphics[width=3in]{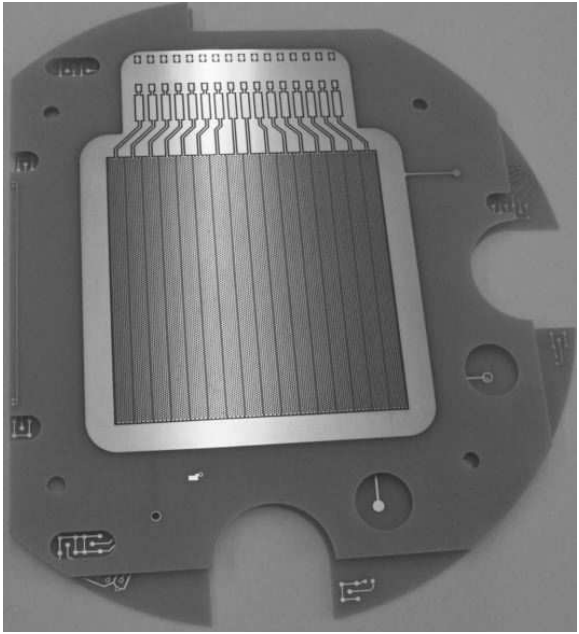}
\caption{Top face of the second LEM stage showing the hole pattern and the subdivision into strips.}
\label{topLEM}
\end{figure}
This suggests
the use of the same formalism of the parallel plate chamber by replacing the gap thickness $d$
by the effective amplification path length within the holes, called $x$, which can be estimated
with electrostatic field calculations as the length of the field plateau along the hole.
The gain is then expressed as  $G_{LEM} = e^{\alpha x}$, where $\alpha$
is the first Townsend coefficient
at the maximum electric field $E$ inside the holes.
For example,  simulations indicate that $x\simeq 1$~mm
for a LEM thickness of 1.6~mm, hence, $\alpha(cm^{-1})=\ln G/(0.1~cm)$.
In order to obtain high
gains ($\sim$1000) in stable conditions we used two multiplication stages.
Ionization electrons undergo multiplication into a first LEM plane and the
resulting charge is then driven into a second LEM plane for further
multiplication.
The amplified charge is readout by measuring two orthogonal coordinates, one using the induced signal 
on the segmented upper electrode of the second LEM itself and the other by collecting the electrons on a segmented anode. 
In this first production both readout planes are segmented with 6~mm wide strips (see Figure~\ref{topLEM}), for a total of 32 readout channels for a $\sim$10x10~cm$^2$ active area.
Transverse segmentations down to 2--3~mm will be tested in the near future.

In the double stage LEM setup, the gain can be symbolically expressed as
\begin{equation}
G \equiv G_{LEM1}G_{LEM2} \simeq e^{\alpha(E_1)x} e^{\alpha(E_2)x}
\end{equation}
where $G_{LEM1}$ (resp. $G_{LEM2}$) are the gains in the first (LEM1)
and second stage (LEM2), $\alpha(E)$ is the first Townsend coefficient
at the maximum electric fields $E_1$ (resp. $E_2$)
inside the holes of LEM1 (resp. LEM2), and $x$ is the effective amplification
path length within the hole.

We have operated the chamber at room temperature and approximately atmospheric pressure (1.2~bar), and
at cryogenic temperature at $\sim$1~bar pressure.
In order to detect cosmic ray tracks, we considered two modes of operation: a ``high'' gain mode at room temperature with $G^{high}\simeq$1000 and
``low'' gain mode at cryogenic temperature with $G^{low}\simeq$10 to compensate for the $\sim$800 times higher dE/dx in LAr.
These modes have been achieved with approximately equal gain on both LEM stages:
\begin{eqnarray}
G^{high} \approx G_{LEM}^2 & \simeq & 30^2 \Leftrightarrow \alpha(E) \simeq 35 cm^{-1}\\ 
G^{low} \approx G_{LEM}^2 & \simeq & 3^2 \Leftrightarrow \alpha(E) \simeq 10 cm^{-1}.
\end{eqnarray}

MAGBOLTZ simulations were used to estimate the required electric fields in the various gas configurations.
At $T=300$~K and $p=1.2$~bar a gain $G\simeq G^{high}$ corresponds to $E\simeq 14$~kV/cm,
while at $T=87$~K and $p=1$~bar a gain $G\simeq G^{low}$ corresponds to $E\simeq 25$~kV/cm.

\section{Design of the 3~lt chamber prototype}
\label{sec:LEM-TPC}
\begin{figure}[htb]
\centering
\includegraphics[width=3in]{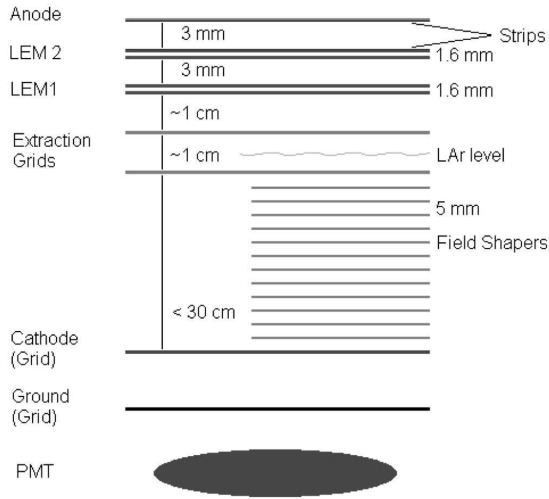}
\caption{Schematic of a LEM-TPC setup showing the LAr level between the two extraction grids. When operated in pure Ar gas, radioactive sources were placed on the ground grid above the PMT.}
\label{LEMScheme}
\end{figure}

\begin{figure}[htb]
\centering
\includegraphics[width=2.8in]{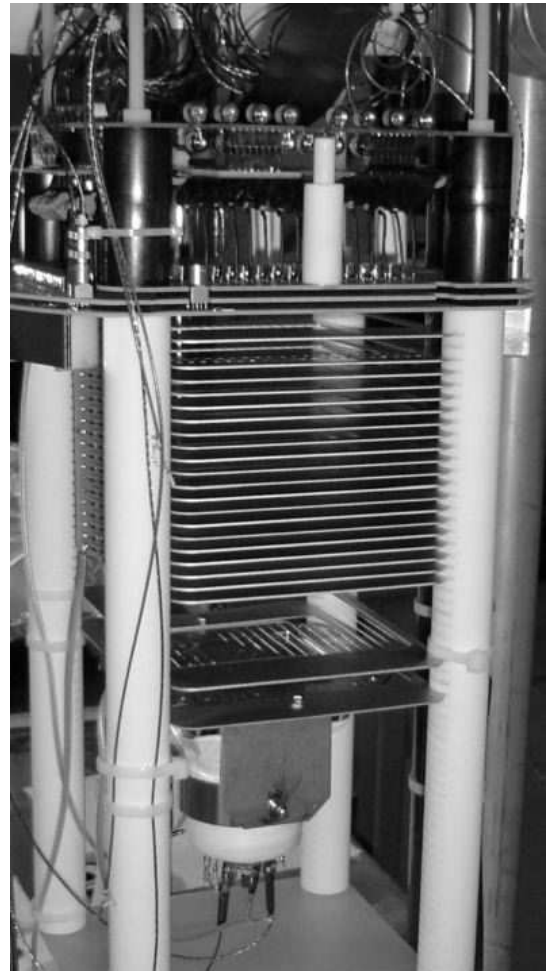}
\caption{Assembly of a LAr~LEM-TPC prototype.}
\label{LEM-TPCSetup}
\end{figure}

We constructed a $\sim$3~lt active volume LAr TPC with a Large Electron Multiplier~(LEM) readout system, as shown 
schematically in Figure~\ref{LEMScheme}. A LAr drift volume of 10x10~cm$^2$ cross section and with an adjustable 
depth of up to 30~cm is followed on top by a double stage LEM positioned in the Ar vapour at about 1.5~cm 
from the liquid. Ionization electrons are drifted upward by a uniform electric field generated by a system of 
field shapers, extracted from the liquid by means of two extraction grids
positioned across the liquid-vapour interface and driven onto the LEM planes.
The extraction grids were constructed as an array of parallel stainless steel wires of 100~$\mu$m diameter with 5~mm spacing.
A cryogenic photomultiplier (Hamamatsu R6237-01) is positioned below the drift region and electrically 
decoupled from the cathode at high voltage by a grid close to the ground potential. The photomultiplier is coated with
tetraphenylbutadiene~(TPB) that acts as wavelength shifter for 128~nm
photons of Ar scintillation. A photograph of the whole setup is shown in Figure~\ref{LEM-TPCSetup}.

Signals from LEM and anode strips are decoupled via high voltage capacitors and routed to a signal
collection plane placed a few centimeter above the anode. Each signal line is equipped with a surge
arrester~\cite{EPCOS} to prevent damaging the preamplifiers in case of discharges.
The detector is housed inside a vacuum tight dewar and Kapton flex-print are used to connect the
signal lines on the signal collection board to the external readout electronics.
The flex-prints exit the dewar through a slot cut in an UHV flange and sealed with a cryogenic
epoxy-resin to maintain vacuum tightness.

The detector is first evacuated down to a residual pressure of a few $10^{-6}$~mbar and then operated in
pure Ar gas at room temperature or in double phase Ar at LAr temperature. When operated in cryogenic environment, the stability of
thermodynamic conditions is ensured by keeping the detector dewar immersed in an external
LAr bath. The internal pressure of the vapour in equilibrium with the liquid is thus the same of the
external atmospheric pressure. The LAr level in the detector dewar is adjusted to be in between the
two extraction grids and it is continuously monitored by three capacitive level meters hanging from the
first LEM plane, with a precision of 0.5~mm.

\begin{figure}[htb]
\centering
\includegraphics[width=2.8in]{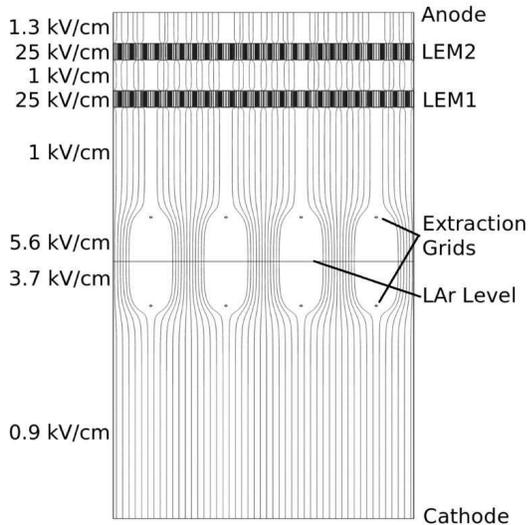}
\caption{Electric field lines in the double phase operation.}
\label{fieldLines}
\end{figure}

The voltages applied to the cathode, field shapers, extraction grids and LEM stages must guarantee the
efficient extraction of the ionization electrons from the liquid, the gain and the transparency of the LEMs for
the drifting electrons. Typical values of the electric fields for operation in gas and in double phase
Ar are shown in Table~\ref{tableField} for gains discussed in section~\ref{sec:LEM}.
\begin{table}[htbp]
\begin{center}
\begin{tabular}{|c|c|c|}
\hline
Fields in (kV/cm)   & Gas operation   & liquid-vapour operation \\
\hline
E Anode-LEM$_2$     & 1               & 1.3                     \\
E LEM$_2$           & 14              & 25                      \\
E LEM$_2$-LEM$_1$   & 0.7             & 1                       \\
E LEM$_1$           & 14              & 25                      \\
E LEM$_1$-Grid      & 0.5             & 1                       \\
Extraction (GAr)    & 0.5             & 5.6                     \\
Extraction (LAr)    & 0.5             & 3.7                     \\
Drift field         & 0.5             & 0.9                     \\
\hline
\end{tabular}
\end{center}
\caption{Typical values of electric fields in kV/cm for operation in GAr at 1.2~bar and in
  double phase conditions at 1~bar.}
\label{tableField}
\end{table}
The Figure~\ref{fieldLines} shows the electric field lines from the
cathode to the anode for the double phase operation. The electric fields are set increasingly from
the drift region towards the anode such that fields lines starting
at the cathode reach the anode (transparency).

\section{Readout electronics}
\label{sec:electronics}
Signals from the LAr~LEM-TPC need first to be pre-amplified and then shaped for noise reduction and 
double track resolution. We felt that it was especially important to develop customized preamplifiers, since they are not 
available commercially and their performance is pushed to the state-of-the-art by the physics requirements, 
making them an essential part of the experimental apparatus.

Our main requirement was a charge sensitivity of $\sim$10~mV/fC with a dynamic range of about 3~V, while keeping a signal/noise ratio of at least 10 for 
a signal of 1~fC and 200~pF input capacitance. Shaping times at the $\mu$s level are adequate for the typical low rate 
application of LAr detectors.

\begin{figure}[htb]
\centering
\includegraphics[width=2.8in]{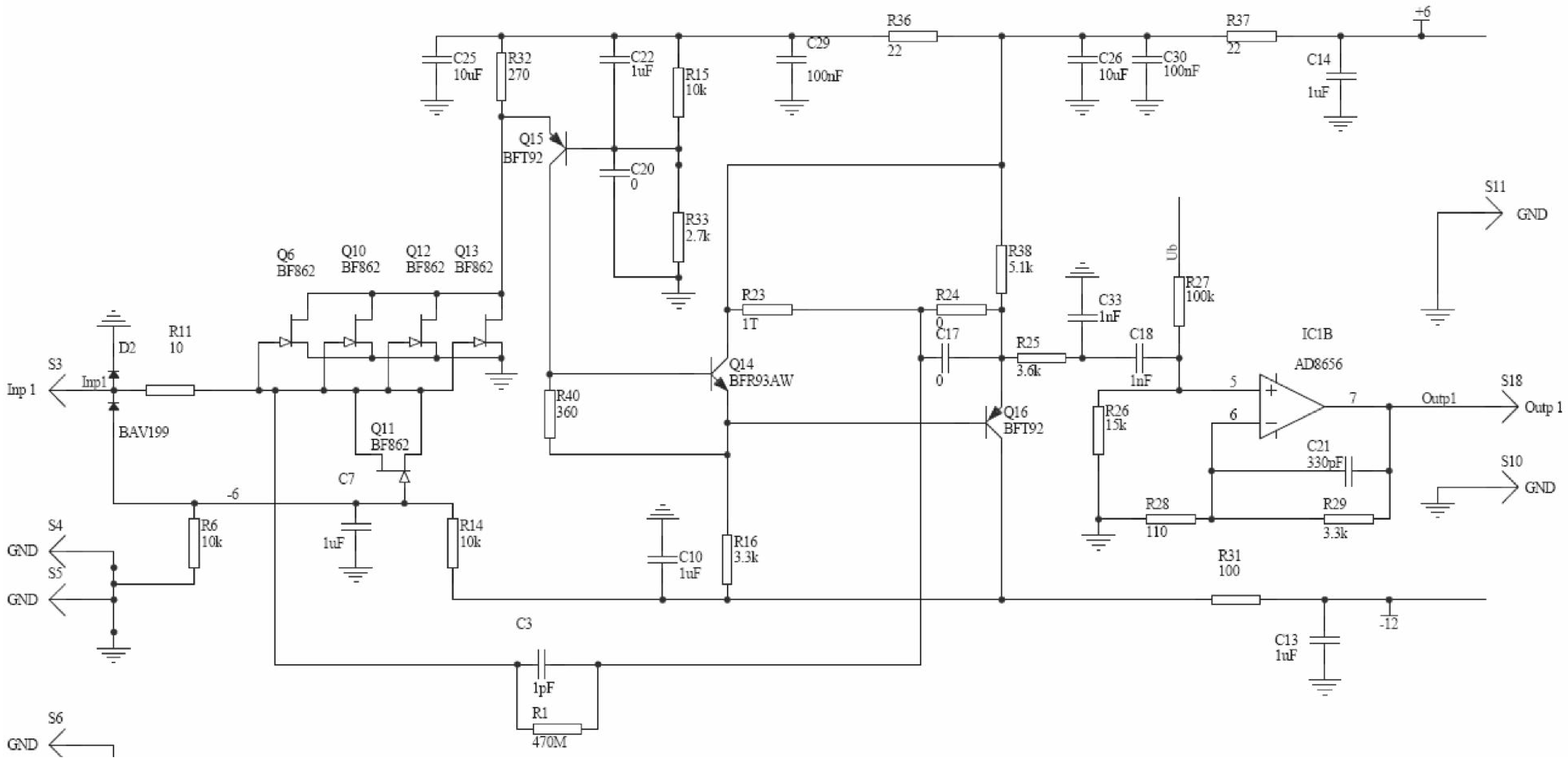}
\caption{Schematic of the preamplifier.}
\label{preampScheme}
\end{figure}

The preamplifier schematic of our design, inspired from~\cite{Boiano04}, is shown in Figure~\ref{preampScheme}.
The charge integrator has four low noise 
BF862~FET transistors from Philips Semiconductor connected in parallel to match a high detector capacitance. 
Its charge sensitivity is 1~mV/fC, as determined by the 1~pF feedback capacitance. The integrator stage is followed 
by a RC-CR shaper with a gain of about 10. The amplifiers were designed to be compatible with both positive and negative 
inputs and are provided with an input for the adjustment of the output baseline, in order to utilize the full dynamic range 
of the digitizer.

\begin{figure}[htb]
\centering
\includegraphics[width=2.8in]{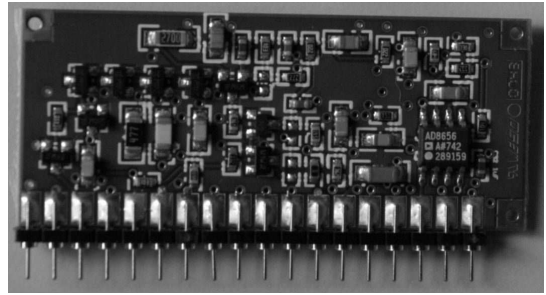}
\caption{Hybrid preamplifier housing two channels.}
\label{preampPic}
\end{figure}

Two preamplifier channels are housed on a single hybrid. An exemplar is shown in Figure~\ref{preampPic}.
While maintaining the same integrator decay time constant of about 500~$\mu$s, different amplifier versions 
were produced with different shaper integration and differentiation time constants.
For this work we used a preamplifier with shaper integration and differentiation time constants of 0.6~$\mu$s and 2~$\mu$s, respectively.
We measured a sensitivity of about 12~mV/fC and a signal to noise ratio of 10 for 1~fC input charge and 200~pF input capacitance.

We felt that it was necessary to cooperate with industry to develop a complete detector readout system for LAr TPCs. 
Therefore, we started a collaboration with CAEN
to develop an industrial version of a complete readout system for LAr TPCs~\cite{CAEN}, which would be compatible with our preamplifiers.
We agreed with CAEN on the main guidelines of the new readout system:
\begin{enumerate}
\item  a single readout board houses 32 preamplifier channels, the corresponding digitizers, the trigger logic and the readout system;
\item  each preamplifier output is digitized by a dedicated 12 bit 2.5 MS/s serial ADC, with no multiplexing;
\item  the system operates as a waveform digitizer: the 32 serial outputs of the ADCs are connected to one FPGA 
which continuously reads the digital samples and writes them, in parallel for all the channels, 
into an array of circular memory buffers;
\item  each channel operates independently from the others and it is triggered when the digitized signal crosses 
a programmable digital threshold;
\item  a trigger defines a time window in which the waveform is acquired, saving a programmable number of samples 
before the trigger and a programmable number of samples after the trigger occurs;
\item  the trigger of each individual channel is propagated to all the other channels of the system, 
even if in different boards or crates, to create a Trigger Alert, which causes the other channels to lower their thresholds, 
in order to be able to trigger on smaller signals.
\end{enumerate}

A prototype of the system was delivered by CAEN in Spring 2008 and it was used for the work described in this paper.

\section{Operation of LEM-TPC in pure GAr at ambient temperature at 1.2~bar}
\label{sec:gas}
\begin{figure}[hbt]
\centering
\includegraphics[width=3in]{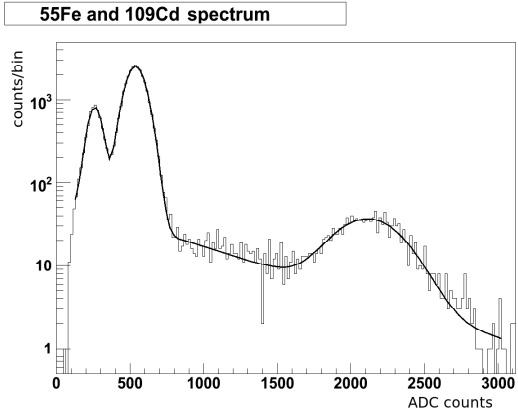}
\caption{Amplitude spectrum from the LEM electrode showing, from left to right, the $^{55}$Fe escape peak, the $^{55}$Fe full energy peak and the $^{109}$Cd peak.
The continuous line is a fit with three independent gaussians and an exponential background.}
\label{LEMAmplitudeSpectrum}
\end{figure}

\begin{figure}[hbt]
\centering
\includegraphics[width=3in]{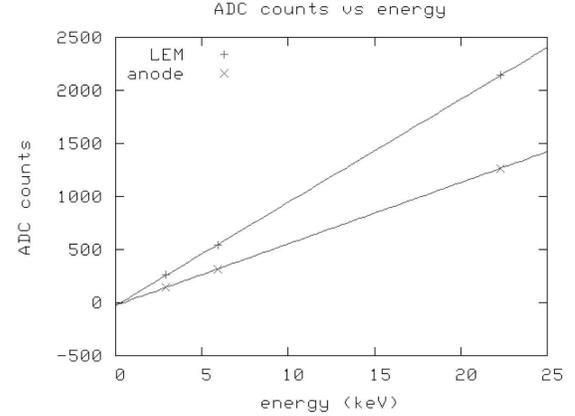}
\caption{Radioactive source peak positions from the LEM and the anode electrodes versus nominal energy depositions.}
\label{energyLinearity}
\end{figure}

The goal of the LEM-TPC operation in Ar gas at ambient temperature was the observation of cosmic muon tracks, requiring a sensitivity down to a few keV per strip.
We chose a double stage LEM gain of $\sim$1000 to give a minimum ionizing signal of the order of hundred ADC counts, well above the noise level.
$^{55}$Fe and $^{109}$Cd radioactive sources of 6.9~kBq and 0.5~kBq, respectively, were inserted in the detector, below the cathode grid, to determine and monitor the gain of the system.
The CAEN prototype readout system was triggered independently on each channel with a threshold of about 1.5~keV.
Source events were selected requiring low hit multiplicity events. The amplitude spectrum from the LEM electrode is shown in Figure~\ref{LEMAmplitudeSpectrum}.
All radioactive sources are clearly visible and the fitted resolutions are given in Table~\ref{tableSpectrum}.
\begin{table}[htbp]
\begin{center}
\begin{tabular}{|c|c|c|c|}
\hline
                     & $^{55}$Fe Escape Peak & $^{55}$Fe Full Peak & $^{109}$Cd \\
\hline
Energy (keV)         & 2.9                   & 5.9                 & 22.3       \\
FWHM Resolution (\%) & 42                    & 29.3                & 24.7       \\
\hline
\end{tabular}
\end{center}
\caption{FWHM energy resolution of $^{55}$Fe and $^{109}$Cd radioactive sources derived from the signal spectrum of the LEM electrode.}
\label{tableSpectrum}
\end{table}
The fitted peak positions, both from the LEM and the anode electrode spectra, are compared to the nominal energy depositions of the radioactive sources in Figure~\ref{energyLinearity},
showing very good linearity.

\begin{figure}[hbt]
\centering
\includegraphics[width=3in]{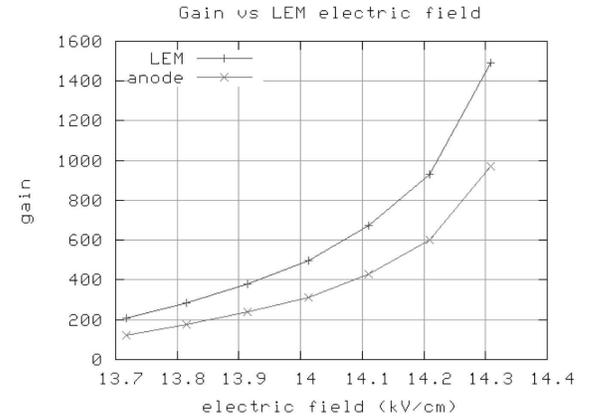}
\caption{Dependence of the gain on the electric field inside the LEM holes from LEM electrode and anode spectra.}
\label{gainVsField}
\end{figure}

Using the $^{109}$Cd peak position we measured the gain for range of the electric field inside the LEM holes
(defined as the ratio of the potential difference across the LEM faces to the LEM thickness) of $14 \pm 0.3$~kV/cm as reported in Figure~\ref{gainVsField},
demonstrating the tunability of the LEM device.

\begin{figure}[t]
\centering
\includegraphics[width=3in]{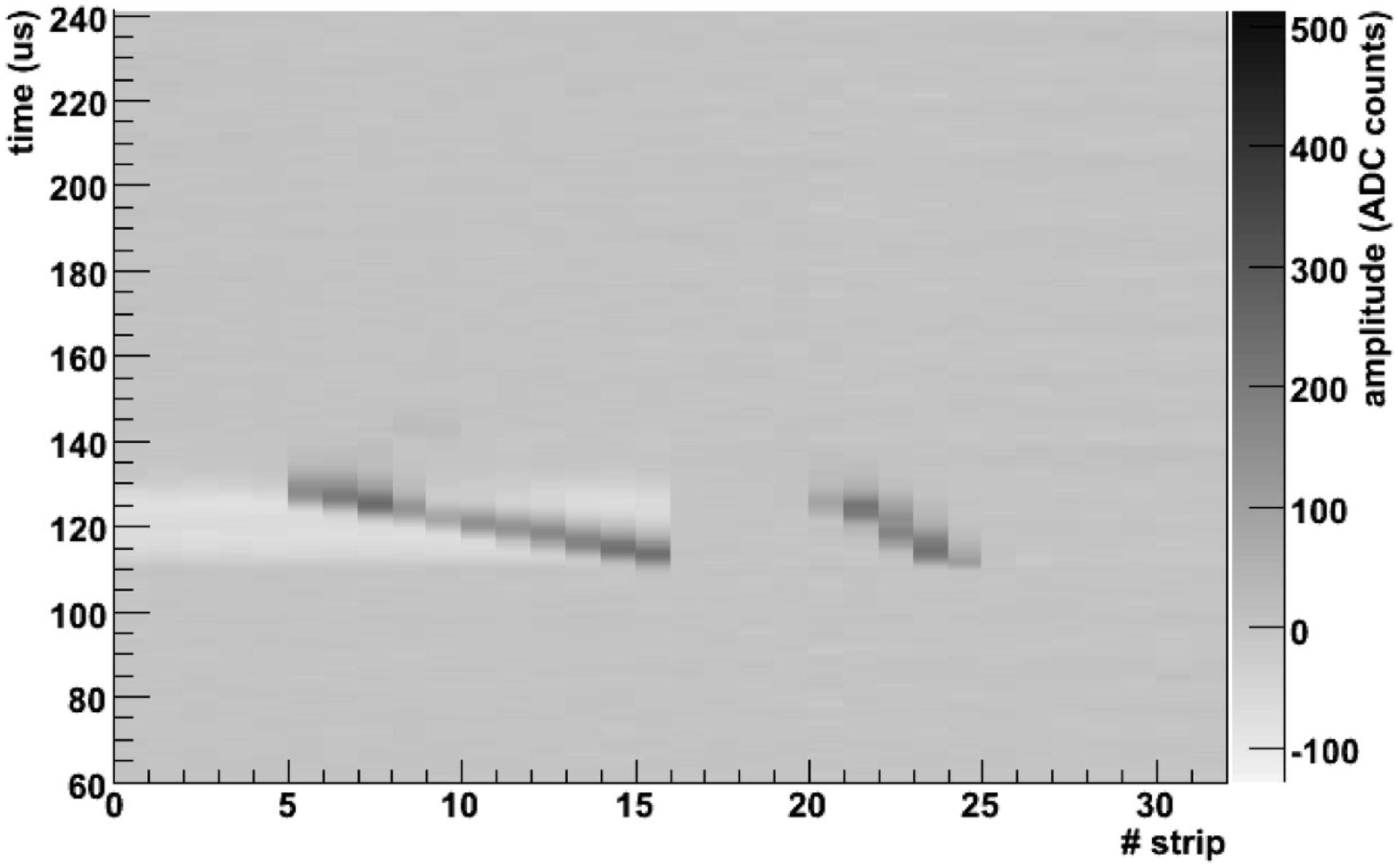}
\includegraphics[width=3in]{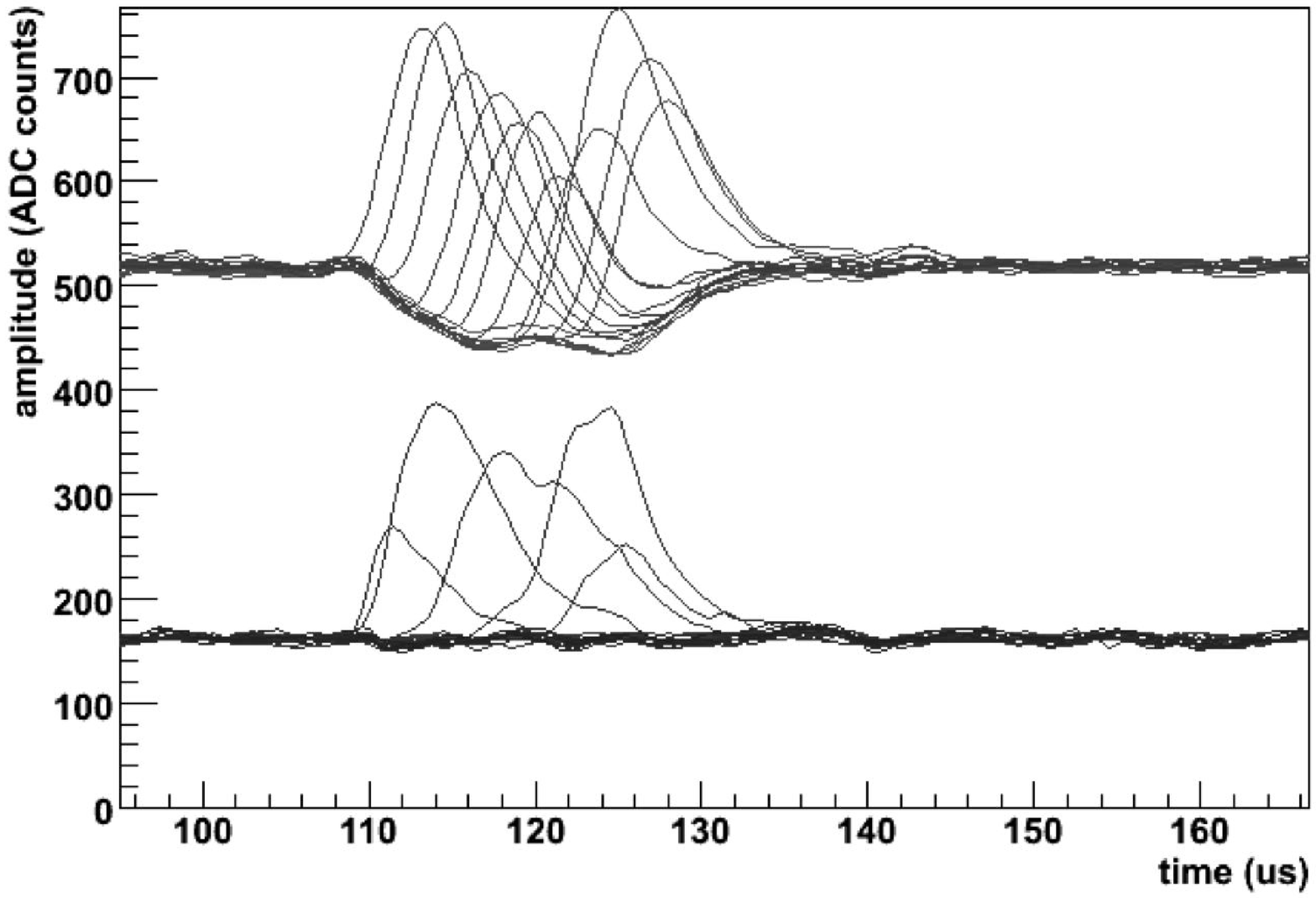}
\caption{Display of a typical cosmic ray event in Ar gas. Channels 0-15 (red upper traces in the bottom picture) are connected to LEM strips and channels 16-31 (blue lower traces in the bottom picture) to anode strips.}
\label{gasEvent}
\end{figure}

High hit multiplicity events showed crossing cosmic muons, as displayed for example in Figure~\ref{gasEvent}.
The top picture shows the arrival time of the signal versus the strip position, both for the LEM and anode electrodes,
allowing the spatial reconstruction of the track. The gray scale on the right is proportional to the signal amplitude.
The bottom picture represents the recorded waveforms for all the channels. An excellent signal to noise ratio is visible.
The base line distortion apparent on the LEM electrode signals is not due a failure or cross-talk of the electronics.
We interpret it as a capacitive pickup on the upper LEM electrode of the physical signals induced on the lower LEM face,
which could be cured by connecting the lower LEM face to a filter capacitor.

\section{Operation of LEM-TPC in double phase Ar}
\label{sec:liquid}
\begin{figure}[t]
\centering
\includegraphics[width=3in]{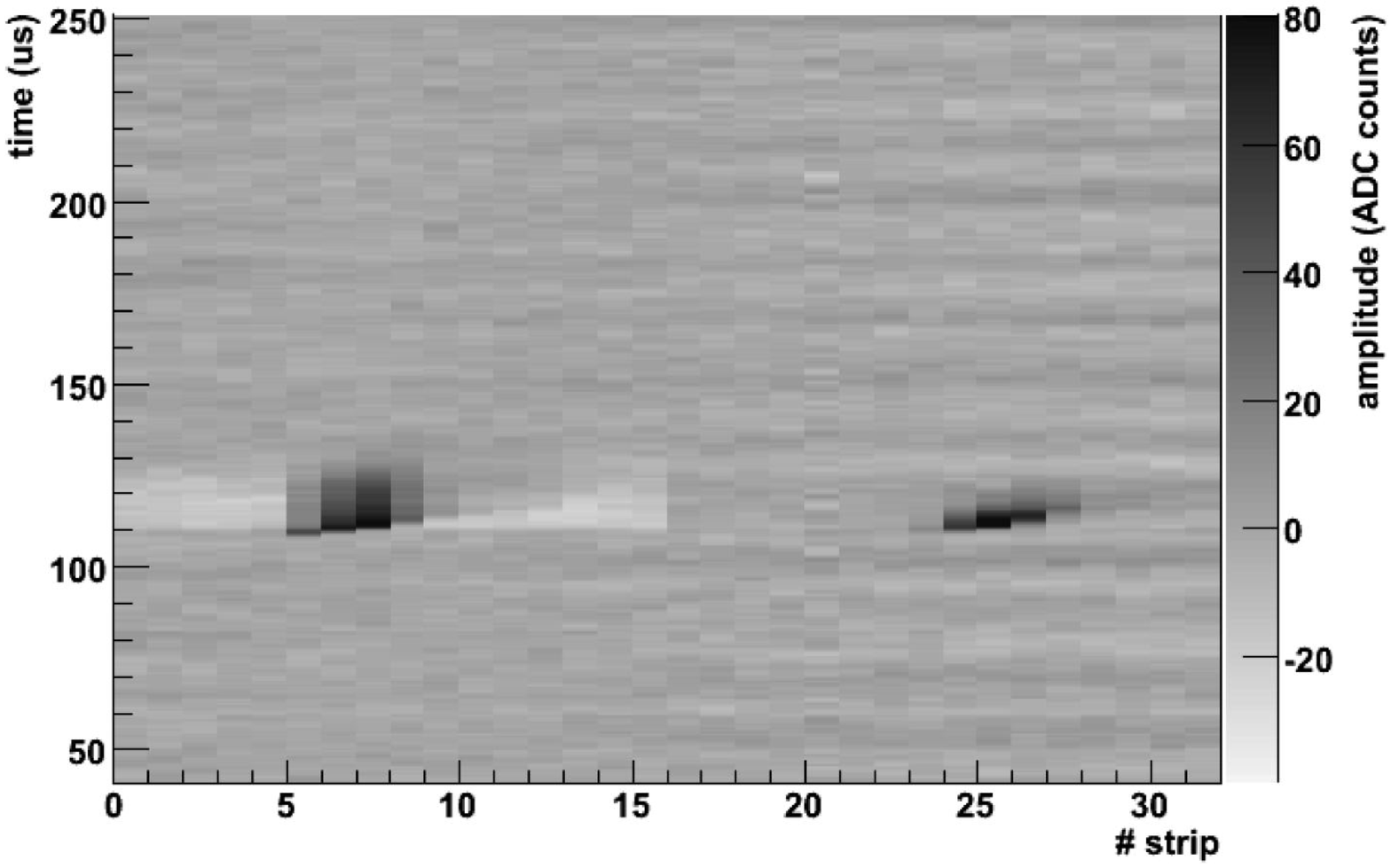}
\includegraphics[width=3in]{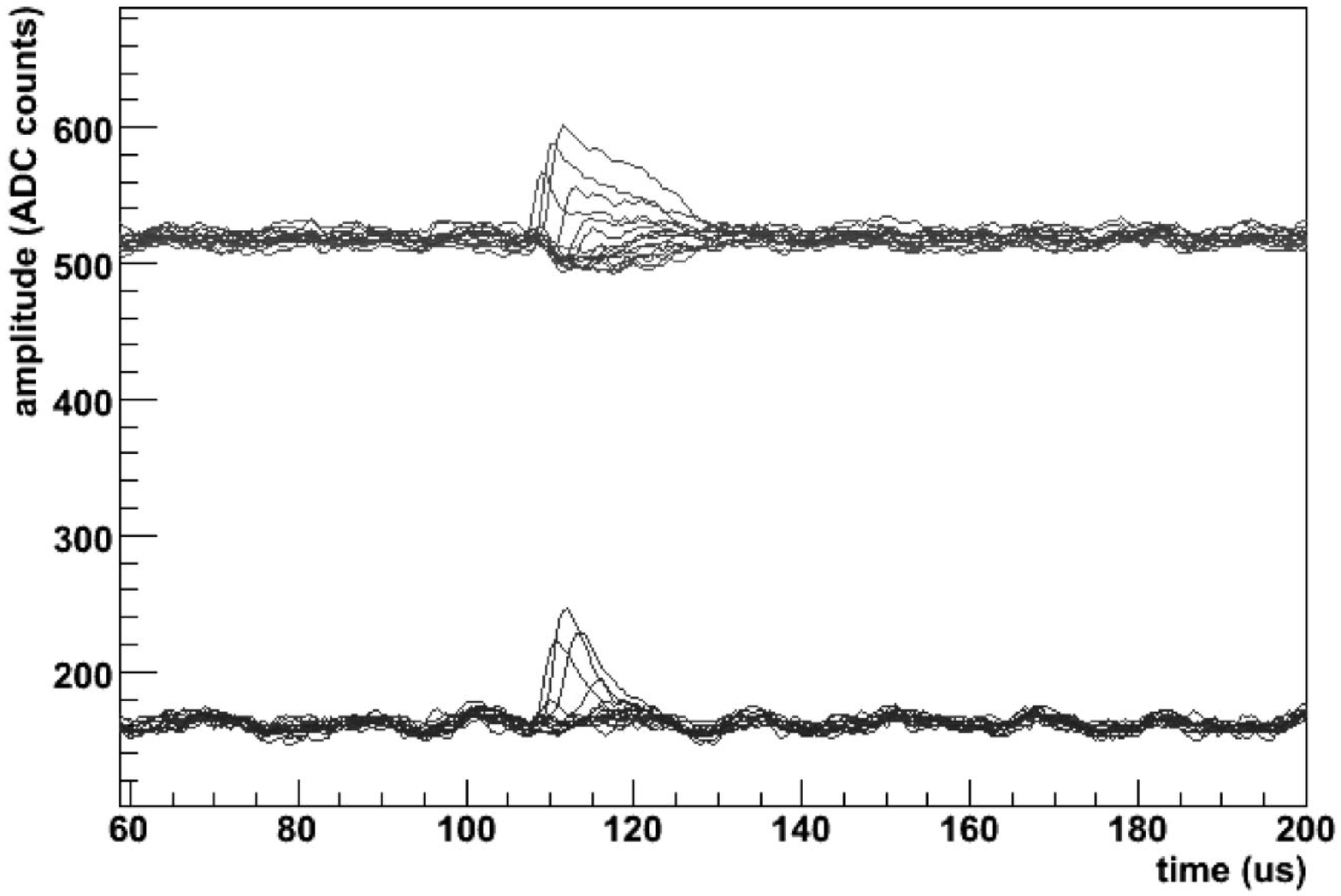}
\caption{Display of a typical cosmic ray event in double phase operation. Channels 0-15 (red upper traces in the bottom picture) are connected to LEM strips and channels 16-31 (blue lower traces in the bottom picture) to anode strips.}
\label{liquidEvent}
\end{figure}

\begin{figure}[hbt]
\centering
\includegraphics[width=3in]{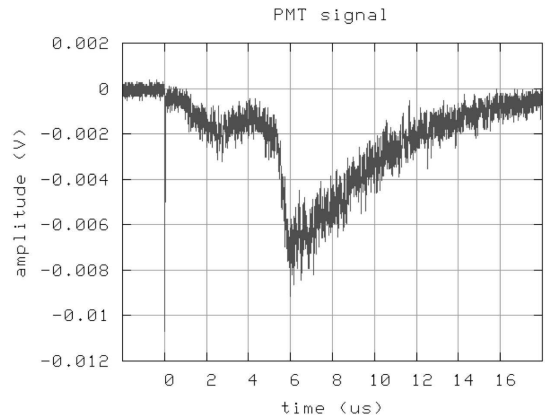}
\caption{Typical signal waveform of the immersed PMT.}
\label{fig:pmtsignal}
\end{figure}

In double phase operation the device gain was set to about 10 to compensate for the higher energy deposition in LAr of cosmic ray muons.
The radioactive sources used at room temperature conditions were removed because they were not suitable for cryogenic operation.
In this mode of operation, the PMT signal proved to be very useful in order to analyse the faith of primary ionization electrons (See Figure~\ref{fig:pmtsignal}): 
a fast light peak indicated the direct scintillation of the crossing cosmic muon. A second light peak, shifted in time if the ionizing
track did not cross the liquid surface, corresponded to the proportional scintillation (luminescence) in the high field region in the
vapour just above the liquid. A third peak was interpreted as light produced during the multiplication avalanche, which escapes the
LEM holes. Indeed, the size of the 3rd light peak was correlated with the electric field inside the LEM.
Right after filling phase, we often observed that this peak was absent, even under a strong LEM electric field. This was
presumably due to liquid argon wetting of the LEM planes. Systematically however, this situation reverted itself to a normal
condition after waiting a few hours. This is evidence that liquid argon does not permanently wet the LEM.
No attempt was made yet to continuously purify the LAr to improve the electron lifetime, still muon tracks were visible during a few days of stable operation.
An example of cosmic muon track is shown in Figure~\ref{liquidEvent}. This represents a proof of principle of the operation of a double phase LAr~LEM-TPC as a tracking device.

\section{Conclusions}
In this paper, we presented for the first time results on the successful operation of a novel kind of liquid argon
Time Projection Chamber based on a finely segmented Large Electron Multiplier (LEM) readout system.
Differently from the ICARUS LAr TPC, the LAr~LEM-TPC is operated in double 
phase (liquid-vapour) pure Ar to allow signal amplification in the gas phase.
In addition, the LEM gives flexibility in the amount of 
multiplication of the primary ionization electrons, thus adapting to a wide range of physics 
requirements.
By finely segmenting the LEM, a bubble-chamber-like image of ionizing events will be obtained, retaining
the salient features of the ICARUS imaging technology, although with a much lower energy
threshold.

We successfully constructed and operated a 3~lt LAr~LEM-TPC.
Cosmic ray tracks have been successfully observed in pure gas at room temperature and
in double phase Ar operation. Radioactive sources have been used
to characterize the gain, resolution and linearity of the system.

This represents a proof of principle of the operation of a double phase LAr~LEM-TPC as a tracking device.
It is an important milestone in the
realization of very large, long drift (cost-effective) LAr detectors
for next generation neutrino physics and proton decay experiments, as
well as for direct search of Dark Matter with imaging devices.

\end{document}